\documentstyle[namedreferences,psfig]{spackap}

\def\gtsima{$\; \buildrel > \over \sim \;$}
\def\gsim{\lower.5ex\hbox{\gtsima}}
\def\ltsima{$\; \buildrel < \over \sim \;$}
\def\lsim{\lower.5ex\hbox{\ltsima}}

\begin{opening}
\title{Measurements of the $^{12}$C/$^{13}$C ratio in Planetary \\ Nebulae
and implications for stellar evolution}

\author{F. \surname{Palla}}
\author{D. \surname{Galli}}
\institute{Osservatorio Astrofisico di Arcetri \\ L.go E. Fermi, 5 - Firenze (Italy)}
\author{R. \surname{Bachiller}}
\author{M.P\'erez \surname{Guti\'errez}}
\institute{Observatorio Astron\'omico Nacional \\ Alcala de Henares (Spain)}

\end{opening}

\runningauthor{F. Palla et al.}
\runningtitle{$^{12}$C/$^{13}$C in Planetary Nebulae}

\begin{document}


\begin{abstract}

We present the results of a study aimed at determining the $^{12}$C/$^{13}$C
ratio in two samples of planetary nebuale (PNe) by means of mm-wave
observations of $^{12}$CO and $^{13}$CO. The first group includes
six PNe which have been observed in the $^3$He$^+$ hyperfine transition;
the other group consists of 23 nebulae with rich molecular envelopes.
We have determined the isotopic ratio
in 14 objects and the results indicate a range of values between 9 and 23.
In particular, three PNe have ratios well below the value predicted by 
standard evolutionary models ($\gsim 20$), indicating that some 
extra-mixing process has occurred in these stars.
We briefly discuss the implications of our results for standard and 
nonstandard stellar nucleosynthesis.

\end{abstract}


\section{Introduction}

In the PN phase, stars more massive than solar
return to the ISM material that has been processed
in the stellar interior. This matter mixes with the surrounding medium
and modifies the original abundances of elements. The contribution
of PNe to the galactic chemical evolution is
particularly important for $^3$He which, together with deuterium (D),
plays a fundamental role in testing the standard Big Bang nucleosynthesis
model. 
While the evolution of D is well understood,
that of $^3$He still encounters serious problems
which cast doubts on the usefulness of this isotope as a test of
Big Bang nucleosynthesis models (e.g. Galli et al. 1995).
In fact, observations of $^3$He toward PNe and HII regions give values
of the abundance that differ by almost two orders of magnitude:
[$^3$He/H]$\sim 10^{-3}$ in PNe and $\sim 10^{-5}$ in HII regions
and in the solar system. However, the abundance in PNe is
exactly that predicted by standard stellar evolution models for stars
of mass 1-1.5 M$_\odot$. The main question is then: if low mass stars
produce a lot of $^3$He and return it to the ISM during the PN--phase,
why don't we see it at a level much higher than observed in HII regions
and the solar system as all standard galactic evolutionary models predict?

Possible solutions to this question have been extensively discussed in this 
Workshop (cf. the reviews by Tosi and by Charbonnel).
The most interesting suggestions invoke the existence 
of nonstandard mixing mechanisms which operate during the red giant phase
of stars with M$_\ast\lsim 2$~ M$_\odot$.
If such mechanisms are indeed at work, an unavoidable
consequence is that the ratio of $^{12}$C/$^{13}$C in PN ejecta should be
much {\it lower} than in the standard case. For a 1 M$_\odot$ star, the
predicted ratio is about 5 against the standard value of 25--30
(Charbonnel 1995; Sackmann \& Boothroyd 1997). 
Therefore, it is very important to have a precise measure
of the isotopic ratios in those PNe where the $^3$He abundance has been
determined. Should these stars show a $^{12}$C/$^{13}$C ratio
close to 25--30, then no modifications to the standard stellar models
would be required. Otherwise, one has to invoke another selective
process (mixing, diffusion etc.) that operates on some isotopes but
not on $^3$He. However, the number of PNe with $^3$He measurements is 
small (see Rood this volume), whereas the suggested physical processes
should be quite general and should affect the nucleosynthetic yields of 
all stars of mass less than $\sim 2$~M$_\odot$. 
Thus, the interest of measuring the carbon isotopic ratio in a sample 
of PNe as large as possible. In this contribution, we present the initial
results of such a study.

\section{Measuring the isotopic ratio from mm-wave observations}

Molecular line observations at mm-wavelengths provide the most powerful 
method to estimate the $^{12}$C/$^{13}$C ratio in PNe. In particular, 
the $^{12}$CO/$^{13}$CO ratio
should faithfully reflect the atomic $^{12}$C/$^{13}$C ratio, since the mechanisms
which could alter the $^{12}$CO/$^{13}$CO ratio are not expected to be at
work in PNe. Namely, (i) the kinetic temperatures in PN
envelopes (25--50 K) is high enough
that the isotopic fractionation should not operate, and (ii) selective
photodissociation is expected to be compensated by the isotope change reaction
$^{12}$CO+$^{13}$C$^+ \rightarrow ^{13}$CO+C$^+$ 
which is faster than the $^{13}$CO
photodestruction in PN envelopes (e.g. Likkel et al. 1988). 
However, although the J=1--0 and J=2--1 lines of $^{12}$CO have 
been extensively observed in PNe (e.g. Huggins et al. 1996), very few 
observations of the $^{13}$CO lines are available and the value of
the isotopic ratio is presently unknown.

Our project consists of two steps. In the first one, we have carried
out high quality observations of $^{12}$CO and $^{13}$CO in six PNe where the 
$^3$He abundance is known from the observations of Rood and collaborators.
In the second step, a larger sample of nebulae with strong $^{12}$CO
line emission has been observed in $^{13}$CO lines in order to determine
the isotopic ratio in PNe {\it without} $^3$He measurements. 
Galli et al. (1997) have argued that 
extra-mixing processes need not to be at work in {\it all} low-mass 
stars in order
to reconcile the predictions of the galactic evolution of $^3$He with
the observational constraints: acceptable results are obtained if about
70\%-80\% of stars with mass lower than 2 M$_\odot$ undergo extra mixing.
This suggestion can be tested by determining the isotopic ratio in a
statistically significant sample of PNe.

\subsection{Results: PNe with $^3$He measurements}

We have observed the six PNe studied by Balser et al. (1997) with the IRAM
30-m telescope in an observing run in November 1996. The observations were 
made in the J=2--1 and J=1--0 lines of $^{12}$CO and $^{13}$CO,
simultaneously. Bachiller et al. (1993) have shown that PNe shells are
characterized by an intrinsic CO (2--1)/(1--0) line ratio in the range
2--5, indicating that the J=2--1 line is more effective for CO searches.
The results of the observations are given in Table~1 with the PN name,
the distance, mass, the $^3$He number abundance 
(given by Balser et al. 1997), the
intensities I$_{21}$ of the J=2--1 lines and the carbon isotopic ratio. 
With the exception of NGC~6720, the values of I$_{21}$ represent upper limits
to the intensity and have been estimated from the line widths deduced
from the expansion velocities listed in the catalog of Acker et al. (1992).

Disappointingly enough, this sample of PNe shows little emission in CO:
the main
and isotopic lines have been firmly detected only in NGC~6720. 
Bachiller et al. (1989) had already detected carbon
monoxide emission in this object and the emission showed a kind of clumpy 
ring, resembling the optical appearance of the nebula. 
We have observed in $^{13}$CO the most prominent 
clumps (three positions) and detected emission in all cases. Thus, the
present data allow to produce a map of the isotopic ratio
across the nebula. The resulting value of $^{12}$C/$^{13}$C=22 is in
agreement with a previous estimate of Bachiller et al.
(1997). Unfortunately, NGC~6720 has an estimated progenitor mass of
$\sim$2 M$_\odot$, at the borderline of the mass range where
the nonstandard mixing mechanism is expected to significantly decrease
the isotopic ratio. Thus, the derived ratio is consistent with both
standard and nonstandard evolutionary models.

We also detected a line around the $^{12}$CO J=1--0 frequency in the central 
position of NGC~6543, but we failed to detect the J=2--1 line at 
relatively low levels. This indicates that the line near the J=1--0
frequency is probably not due to CO. It is interesting to recall 
that the H38$\alpha$ 
recombination line is only separated by 3 MHz (7.8 km/s) from
the $^{12}$CO J=1--0 line. As discussed in Bachiller et al. 
(1992), the H38$\alpha$
line can dominate the emission around the $^{12}$CO J=1--0 frequency in some
nebulae with little or no molecular gas. We believe that this is the
case in the central position of NGC~6543.

Following the discussion of Rood (this volume), the best object for 
testing the
mixing hypothesis is NGC~3242 where the $^3$He measurements are the most 
reliable ones {\it and} the progenitor mass is sufficiently small ($\sim$1.2
M$_\odot$) that the predicted isotopic ratio should be about 10 (instead of 
$\sim$30). However, the excellent signal to noise ratio of our 
observations and the fact that
we searched for CO emission across the whole circumstellar shell imply that
CO is really absent in the nebula. Therefore, we conclude that the isotopic 
ratio cannot be determined in this important object with millimeter line 
observations.


\begin{figure}
\centerline{\psfig{file=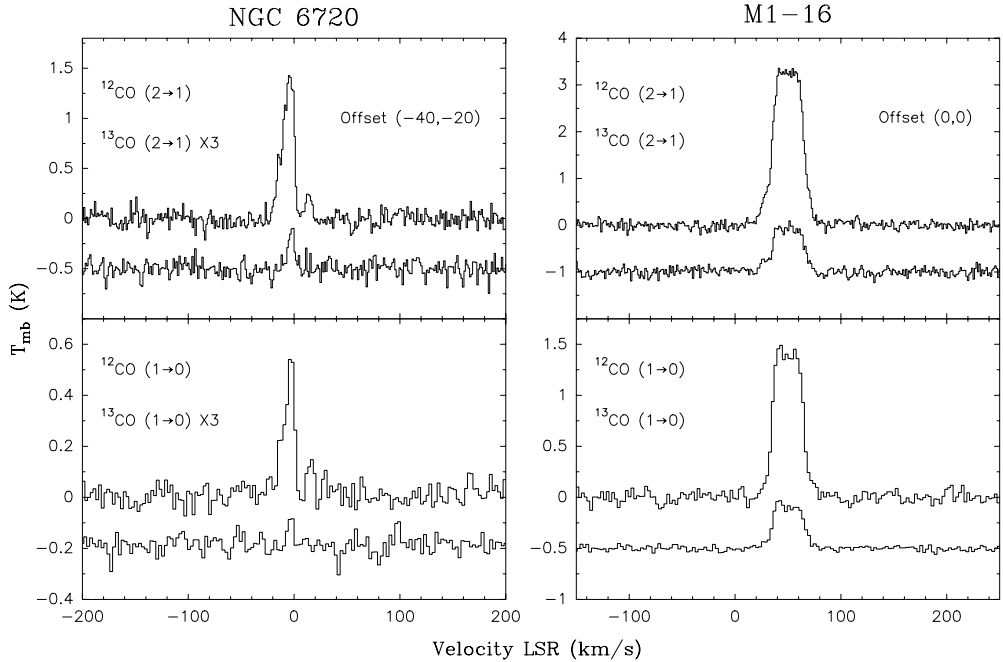,width=14cm,angle=-90}}
\caption{CO emission in PNe. Two examples are shown from the sample of PNe
with ({\it left}) and without ({\it right}) $^3$He measurements. 
The offset in arcsec from the central PN is also indicated.
The stronger J=2--1 transitions are shown in the upper panels, while the 
J=1--0 transitions in the bottom panels. The isotopic lines are detected
in all cases.}
\end{figure}

\begin{table}
\caption[]{Distances, Progenitor Masses and Abundances}
\begin{tabular}{lrrrrrr}\hline
 PN name  & D & M & $^3{\rm He}/{\rm H}$ & I$_{21}$($^{12}$CO) &  
	I$_{21}$($^{13}$CO) & $^{12}$C/$^{13}$C \\
 & (kpc)  &  (M$_\odot$) & ($\times 10^{-3}$) & (K km s$^{-1}$) & 
	(K km s$^{-1}$) & \\
\hline
IC~289   & 1.43 & $<$1.6        & $<$6.7$\pm$2.6     & 0.90 &  &  \\
NGC~3242 & 0.88 & 1.2 $\pm$ 0.2 & 0.92$\pm$0.18      & 0.16 &  &  \\
NGC~6543 & 0.98 & 1.6 $\pm$ 0.2 & $\le$0.49$\pm$0.27 & 0.19 &  &  \\
NGC~6720 & 0.87 & 2.2 $\pm$ 0.6 & $<$0.54$\pm$0.29   & 20.0 & .9 & 22\\
NGC~7009 & 1.2  & 1.4 $\pm$ 0.2 & $\le$0.80$\pm$0.31 & 0.12 &  &  \\
NGC~7662 & 1.16 & 1.2           & $<$0.22            & 0.12 &  &  \\
\hline
\end{tabular}
\end{table}

\subsection{Results: PNe without $^3$He measurements}
We have searched for $^{13}$CO emission in 22 PNe 
using the IRAM 30-m telescope in May 1997. The objects were selected
on the basis of their strong $^{12}$CO emission from the list of
Huggins et al. (1996). The results are: 13 detections, 6 tentative
detections, and 3 upper limits. The parameters of the detected sources
together with the derived isotopic ratios are listed in Table 2.

In order to estimate the $^{12}$CO/$^{13}$CO isotopic ratio, one needs to
make a number of approximations. First, we assume that the
emitting regions fill the antenna beams in the lines of both molecules, 
or that the filling factor is the same (in the case of an extended clumpy
medium). Second, we assume that the rotational levels are thermalized at
a representative uniform temperature of 25 K (see e.g. Bachiller
et al. 1997). Thermalization is indeed a reasonable assumption
for $^{12}$CO and $^{13}$CO, since the dipole moment is very low (about 0.1 Debye). 
Third, if we assume that the emission is optically thin for both
the $^{12}$CO and $^{13}$CO lines, then the $^{12}$CO/$^{13}$CO column density 
ratio is given by the ratio of the integrated intesities. 

The last assumption is likely to be accurate in the case of evolved nebulae 
like the Ring, the Helix, NGC~2346, etc. In fact, Large-Velocity-Gradient
(LVG) models confirm that the $^{12}$CO and $^{13}$CO line emission
is optically thin in these cases. On the other hand, the same assumption 
is not appropriate for the CO lines in young
objects such as CRL~2688, CRL~618, NGC~7027, and M~1-16. In such cases, the 
derived CO column densities and $^{12}$CO/$^{13}$CO column density ratios represent
only crude lower limits.

Finally, one could have weak emission arising from
small optically thick clumps very diluted within the $^{12}$CO and $^{13}$CO beams.
This could be the case in some compact CO envelopes such as the
Butterfly nebula, M~2-9. The CO in M~2-9 is
concentrated in an expanding clumpy ring which has a mean diameter of
6 arcsec (Zweigle et al. 1997). Individual 
clumps in this ring have sizes $<4$ arcsec. The weakness of
the $^{12}$CO and $^{13}$CO lines we observe could be due to the important dilution
of such small clumps within the 30-m beam. The clumps could be optically
thick in CO, and the reported $^{12}$CO/$^{13}$CO intensity ratio would just 
represent a lower limit to the abundance ratio.

In summary, we believe that the isotopic ratios reported here are
reasonably robust estimates for the extended evolved nebulae, namely
NGC~6720, NGC~2346, NGC~7293, NGC~6781, M~4-9, M~2-51, and very likely for
M~1-17 and IC~5117. The values of the isotopic ratio in these nebulae 
are in the range 9 to 23 (within a factor of less than 2), significantly 
lower than the Solar System value of 89. 
The reported values are however nearly within the range 
12 to 36 found in the envelopes of massive AGB stars 
(Greaves \& Holland 1997), which could indicate that the progenitors of 
such nebulae were relatively massive.
Unlike the PNe discussed in the previous subsection, we have not yet 
obtained estimates of the progenitor mass for these stars and therefore
we cannot say if the three PNe with ratios definetely below 20 originated
from low-mass stars. These are the most promising objects for testing 
the hypothesis
of nonstandard burning processes in the AGB phase and work is in progress
to obtain mass estimates (this part is done in collaboration with 
L. Stanghellini and M. Tosi). Similarly, they are prime targets for future
measurements of $^3$He abundances.

\section{Discussion}

The method of using mm-wave transitions is not the only one adequate for
measuring the isotopic abundance in PNe. Clegg (1985) first pointed out the
possibility of using the CIII] multiplet near 1908 \AA~ to obtain a direct
estimate of the isotopic ratio in the {\it ionized} gas of PNe. Very
recently, Clegg et al. (1997) have successfully detected the extremely
weak isotopic line $^{13}$C $^3_{1/2}$P$^\circ_0$-$^1_{1/2}$S$_0$ 
in two PNe (plus a tentative detection in a third one), 
using the HST Goddard High Resolution Spectrograph. 
The $^{12}$C/$^{13}$C ratio has been measured to be 15$\pm$3 and 
21$\pm$11, respectively. In either case, no measurements at mm-wavelenghts
have been made to independently check the derived values. However, as Clegg
et al. point out, the two types of measurements are complementary since
they cannot be performed on the same objects: the UV transitions require
high excitation conditions, while the opposite is true for the mm-wave
lines studied by us.

What are the implications of our results for the understanding of stellar
nuclosynthesis? The fact that the majority of the PNe shows an isotopic
ratio of $\sim$20 implies a moderate degree of depletion from
the constant value achieved during the first dredge-up in the red giant
branch. Such constant value depends on both mass and metallicity of the stars
and varies between 30 for 1 M$_\odot$ and 23 for 2 M$_\odot$ at Z=0.02.
Thus, our results indicate that the observed PNe had rather massive 
progenitors. On the other hand, we also find some PNe with an isotopic
ratio below this constant value, approaching the equilibrium value in the
CN cycle, namely 4-9. Similar low values have been measured in field
population II stars and in globular cluster giants and have provided the
motivation to introduce some extra-mixing process in the standard evolution
(e.g. Charbonnel 1995). 
It will be interesting to try to estimate the
progenitor mass of these PNe since nonstandard mechanisms are effective
only for stars less massive than $\sim 2$~M$_\odot$. These stars should
also have very low $^3$He abundances, and future observations 
should address this issue.
Finally, we remind that in order to reconcile the $^3$He abundances with
galactic chemical evolution models, the majority ($\gsim$70\%) of the
PNe should show an isotopic ratio well below the standard value.
It is interesting to note that 3 out of the 7 PNe with reliable ratios
have values well below 20. However, 
the small number statistics does not allow us to draw any 
conclusion from this preliminary study, but clearly calls for more radio 
observations on a larger sample of PNe. 

As emphasized by Rood, the key object to test theories is NGC~3242.
The lack of a molecular envelope around NGC~3242 can be a 
consequence of the low-mass of the progenitor. It is well known
that the transition from the AGB phase to the PN stage is slow for low-mass
stars and therefore the gas remains exposed to ionizing and dissociating
radiation for a longer time than in the case of more massive stars. 
This is also consistent with the fact that NGC~6720 has a rich molecular
envelope and is the most massive objects of our sample. Therefore, the 
lesson is clear: there
is a trade-off between the need to go to low-mass PNe in order 
to discriminate between
theoretical models and the higher detection rate of molecular envelopes around
more massive objects. Thus, the determination of isotopic ratios may be
a tricky business, but future observations should take up such a challenge.

\begin{table}
\caption[]{Distance, CO Intensity and Isotopic Ratio}
\begin{tabular}{lrrrr}\hline
 PN name  & D & I$_{21}$($^{12}$CO) & 
 I$_{21}$($^{13}$CO) & $^{12}$C/$^{13}$C \\
 & (kpc)  & (K km s$^{-1}$) & (K km s$^{-1}$) & \\
\hline
NGC~7027    & 0.70 & 448.7  & 17.3 & $>$25 \\
NGC~2346    & 0.80 & 14.2   & 0.58 & 23    \\
NGC~7293    & 0.16 & 17.5   & 1.8   & 9.3   \\
NGC~6781    & 0.70 & 28.4   & 1.72  & 20    \\
M~1-16      & 5.45 & 109.0    & 32.3  & $>$3  \\
M~1-17      & 7.36 & 66.2   & 3.2   & 22    \\
M~2-9       & 1.0  & 3.3    & 2.0   & $>$2  \\
M~2-51      & 1.92 & 33.1   & 2.3   & 15    \\
M~4-9       & 1.8  & 32.2   & 1.8   & 18    \\
CRL~2688    &      & 312.5  & 112   & $>$3  \\
CRL~618     & 1.80 & 182.2  & 41.2  & $>$4  \\
OH~09$+$1.3 & 8.0  & 5.0    & 2.5   & $>2$  \\
IC~5117     & 2.10 & 19.6   & 1.4   & 14    \\
\hline
\end{tabular}
\end{table}
\section{Acknowledgements}~It is a pleasure to thank R. Rood, T. Bania and 
D. Balser for numerous and fruitful discussions on the $^3$He measurements. 
This work is done in collaboration with M. Tosi and L. Stanghellini.

{}

\end{document}